\documentclass[12pt]{article}
\usepackage{amsmath}
\usepackage{graphicx,psfrag,epsf}
\usepackage{enumerate}
\usepackage{natbib}
\usepackage{url} 

\usepackage{amssymb}
\usepackage{amsthm}
\usepackage{caption}
\usepackage{algorithm}
\usepackage{algpseudocode}

\newtheorem{theorem}{Theorem}
\newtheorem{lemma}{Lemma}
\newtheorem*{theorem*}{Theorem}
\newtheorem*{lemma*}{Lemma}

\theoremstyle{definition}
\newtheorem{definition}{Definition}

\newtheorem{remark}{Remark}
\newtheorem*{remark*}{Remark}

\usepackage[noabbrev,capitalise]{cleveref}
\usepackage{booktabs}
\usepackage{relsize}
\usepackage{setspace}
\usepackage{makecell}
\usepackage{multirow}

\newcommand{\ignoreme}[1]{}

\newcommand{\SDR}{{SDR}}
\newcommand{\SDP}{{SDP}}
\newcommand{\IDR}{{IDR}}

\newcommand{\captionheading}[1]{#1}
\newcommand{\truprop}{\ensuremath{\theta}}

\newcommand{\LonekA}{L1000A}
\newcommand{\LonekB}{L1000B}

\newcommand{\figwidth}{5.5in}

\newcommand{\Prsymbol}{\mathbb{P}}

\usepackage{todonotes}
\usepackage{xargs} 
\newcommandx{\jacksontodo}[2][1=]{\todo[linecolor=red,backgroundcolor=red!25,bordercolor=red,#1]{#2}}

\usepackage{silence}
\newcommand{\blind}{1}

\begin{document}

\def\spacingset#1{\renewcommand{\baselinestretch}%
{#1}\small\normalsize} \spacingset{1}


\if1\blind
{
  \title{\bf Model-free Sign Estimation for High-Throughput Screenings%
}
  \author{Jackson Loper\hspace{.2cm}\\
    Department of Statistics, University of Michigan\\\\
    Jeffrey Regier \\
    Department of Statistics, University of Michigan}
  \maketitle
} \fi

\if0\blind
{
  \bigskip
  \bigskip
  \bigskip
  \begin{center}
    {\LARGE\bf Model-free Sign Estimation for High-Throughput Screenings%
}
\end{center}
  \medskip
} \fi

\bigskip
\begin{abstract}
In high-throughput screenings, it is common to estimate the effects of many treatments using a small number of independent trials of each. Because little is known about the distributional properties of the measurements from these trials, it is challenging to identify plausible assumptions that can serve as a basis for inferential statistics in this setting.  In this article, we develop a method based on minimal assumptions to infer signs of treatment effects (positive or negative).  The proposed method controls the number of misestimated signs by using the number of sign disagreements between measurements of the same treatment as a proxy for the number of sign errors. In simulations, the proposed method compares favorably with the Benjamini-Hochberg procedure applied to invalid $p$-values, which is currently considered best practice for many high-throughput screenings. For real data from the L1000 cell-perturbation platform, the proposed method outperforms existing practices, which fail to control error at the nominal level in some cases and are needlessly conservative in others.

\end{abstract}

\noindent%
 \vfill

\newpage
\spacingset{1.9} 

\section{Introduction}

Let $\boldsymbol{\theta}$ be a vector of $n$ parameters representing treatment effects. We aim to estimate $\boldsymbol{\theta}$ using $R$ independent measurements of each parameter. Denote the measurements $X^{(1)}_{i},\ldots, X^{(R)}_{i}$ for $i \in \{1,\ldots, n\}$.  If $R$ is large, classic results such as the central limit theorem can be applied to construct confidence intervals for each $\theta_i$.  However, if limited resources are allocated to measure many parameters poorly instead of a few parameters well, as is generally the case in high-throughput screenings \citep{subramanian2017next,srivatsan2020massively,schmidt2022crispr}, then often $R$ is small (e.g., two or three) and there is little basis for assessing the accuracy of each measurement.  Without additional assumptions relating the measurements and the parameters, statistical inference is infeasible in this setting.  

Literature has already been devoted to this small-$R$ regime (\cref{sec:related}), especially on testing hypotheses of no effect.  However, even if a hypothesis of no effect can be rejected, the scientific significance of the effect remains unclear unless the direction of the effect---whether the treatment increased or decreased the response---can be inferred.

In this article, we seek to estimate the sign of each parameter $\theta_i$, denoted $\mathrm{sign}(\theta_i)$, under minimal assumptions about the measurements.  Accurate sign estimation for every parameter may not be feasible, so instead we seek to select a subset of parameters for which the signs can be reliably estimated.  In keeping with the terminology of the false discovery literature, we refer to each parameter in the selected subset as a ``discovery.''  We refer to the misestimation of a parameter's sign as a type S error \citep{gelman2000type}.  

We propose a new method, called Sort, Select, and Estimate (SSE), that identifies a subset of parameters whose signs can be reliably estimated and estimates the sign of each (\cref{sec:method}).  Under minimal assumptions, the proposed method controls the type S error proportion of the estimates, that is, the number of type S errors divided by the total number of parameters estimated (\cref{sec:analysis}).  This form of control is known as ``directional false discovery control'' \citep{benjamini_false_2005}.  This control is guaranteed in the asymptotic limit of large $n$ and holds for any fixed value of $R$.  Guarantees are based on the following central assumption about the measurements: there exists $q \in (0,1)$ such that for every measurement $r$ of any parameter $i$,
$\mathbb{P}(\mathrm{sign}(\theta_{i})=\mathrm{sign}(X_{i}^{(r)}))\geq q$.  We refer to measurements that satisfy this assumption as $q$-faithful. Our results hold for arbitrary $q$, even if $q<0.5$, which implies that some measurements may estimate the sign incorrectly more often than not.  Smaller values of $q$ constitute weaker assumptions and result in fewer discoveries by the proposed method, whereas larger values constitute strong assumptions and result in more discoveries.  In our simulations and case studies, we assume $1/2$-faithful measurements.  This is typically a mild assumption: each sign estimator is no worse than a fair coin toss.  

We use simulations to probe the strengths and weaknesses of the proposed method relative to existing methods (\cref{sec:simulations}).  The new method controls type S error for various noise distributions, consistently holding the error rate below the target level.  Compared to existing methods such as a misspecified parametric model with Benjamini-Hochberg correction (Misspec-BH), the proposed method demonstrates superiority in challenging, high-noise scenarios.  Although Misspec-BH can make more discoveries when its (unwarranted) assumptions hold, it suffers from inflated error rates as noise increases and model assumptions are violated.  The proposed method maintains error control throughout.

We then investigate the performance of the proposed method in the context of high-throughput cell-perturbation experiments from the L1000 cell-perturbation platform \citep{subramanian2017next} (\cref{sec:casestudies}).  This experimental setting is a prime example of high-throughput screenings in which sign estimation is of paramount importance \citep{subramanian2017next,srivatsan2020massively,schmidt2022crispr}. In this setting, each parameter $\theta_{i}$ characterizes the way a particular treatment changes the expression of a particular gene in a population of cells.  The sign of each $\theta_i$ is important to estimate because the gene control circuits that govern cell behavior are defined in terms of upregulation and downregulation of genes \citep{davidson2001genomic}. Several independent experiments are performed to measure each parameter. The variability in each measurement arises from many sources, including batch effects (e.g., variability in lab technician proficiency, humidity) and well effects (e.g., cells housed together in a given well may interact).  These sources are difficult to characterize and have a large impact.  For example, even slight variability in humidity levels can introduce substantial variability in the estimates \citep{stein2015removing}.  Because these sources are poorly understood, it is problematic to apply traditional statistical methods to this data.  

Using the proposed method, which makes few assumptions, we produce error-controlled estimates of the signs of the perturbation effects.  For comparison, we consider current standard practice, which involves applying BH to potentially invalid $p$-values. We find that this procedure based on the BH method yields fewer discoveries than the proposed method; we conjecture that the BH method performs poorly because the $p$-values are invalid.  

We also used the proposed method to compare two different techniques for preprocessing the data from L1000 experiments.  Comparison of different preprocessing methods is an essential task in experimental design, but it is challenging for cell-perturbation experiments in which the variability is so poorly understood.  We show how the assumption of $q$-faithful measurements
can facilitate such comparisons.  The comparisons yield compelling evidence for the superiority of one of the preprocessing methods.

We conclude by identifying three limitations of the proposed method for high-throughput screenings, and suggesting possible future directions to remedy these limitations (\cref{sec:discussion}).

\section{Related work}

\label{sec:related}

Methods have been developed to estimate many parameters using a small collection of independent measurements for each parameter.  However, there are no existing methods that provably control type S error with minimal assumptions in this setting.

\citet{dai2023false} test the null hypothesis of no effect for each parameter under the assumption that the distribution of each measurement is symmetric about the origin under the null hypothesis.  One could use this method to test all null hypotheses of no effect and then use the measurements to estimate the sign for each rejected hypothesis.  However, such an approach can badly fail to control the type S error proportion (cf.~\cref{apdx:mirrorproblems}).

The knockoff framework is another approach that enables hypothesis testing with minimal assumptions \citep{knockoffs}.  Knockoff methods select a subset of regressors that are important in predicting a response.  Knockoffs can be used to control type S errors \citep{cao2023split} in a regression setting.  Our problem setting is different: we are not seeking to estimate the relation of many regressors to a response, nor to test hypotheses about conditional independence.  Instead, we estimate the effect of each treatment in isolation.

\citet{li2011measuring} develop a Bayesian method for identifying salient parameters from measurements under the assumption that the dependence structure between measurements can be modeled using a Gaussian mixture copula with two components.  This method is effective when the distribution of true parameter magnitudes is bimodal, with most parameters having negligible values and a few parameters having large values.  In this case, a Gaussian mixture can accurately identify measurements arising from parameters with large values.  This method is not designed for sign estimation, but we invent a similar method suited for sign estimation.  We compare this approach with the proposed method in \cref{subsec:sims:IIIIII}.

More extensive assumptions have enabled rigorous type S error control. \citet{yu2019adaptive} control type S error by assuming Gaussian noise with fixed variance.   \citet{benjamini_false_2005} control type S error under the assumptions that: (i) the distribution of the measurements is known exactly under a null hypothesis of no effect and (ii) the distribution of each measurement increases stochastically with the value of the parameter.  This latter approach has become the most widely used method for sign estimation in high-throughput cell-perturbation data.

However, cell-perturbation experimentalists often apply the method of \citet{benjamini_false_2005} in cases where its assumptions do not hold.  For example, \citet{srivatsan2020massively} and \citet{schmidt2022crispr} apply the method of \citet{benjamini_false_2005}, with a Mann-Whitney-Wilcoxon (MWW) approach used to produce statistics with known distribution under the null.  However, their study designs do not produce the independent samples required by the MWW test.  Instead, rank statistics are obtained from dependent measurements of many cells that were all housed in the the same well.  Well-sharing induces dependency between the cells, but no correction is made for the dependency.  Indeed, there is little basis for correcting for this dependency as it arises from subtle batch, donor, and well effects.  Nevertheless, this use of the MWW test is considered a best practice \citep{li2022exaggerated}.  A primary motivation of the present work is to design an alternative approach to this misapplication of MWW: an approach that controls type S error under weaker assumptions.

\ignoreme{
 __  __ ______ _______ _    _  ____  _____
|  \/  |  ____|__   __| |  | |/ __ \|  __ \
| \  / | |__     | |  | |__| | |  | | |  | |
| |\/| |  __|    | |  |  __  | |  | | |  | |
| |  | | |____   | |  | |  | | |__| | |__| |
|_|  |_|______|  |_|  |_|  |_|\____/|_____/

}

\section{The SSE method}
\label{sec:method}

\begin{algorithm}[t]
\caption{Sort, Select, Estimate (SSE).}
\label{alg:SSE}
\begin{algorithmic}[1]
\Statex \textbf{Input:} Proposal measurements ($\boldsymbol{X}^{(1)}\in \mathbb{R}^n$), validation measurements ($\boldsymbol{X}^{(2)}\in \mathbb{R}^n$), target type S error proportion ($\beta \in [0,1]$), faithfulness ($q\in [0,1]$) \vspace{.1in}
    \State $\rho \gets \mathrm{argsort}(|X_1^{(1)}|,\ldots,|X_n^{(1)}|)$ \Comment{Sort parameters in descending order}
    \State $U_k \gets |\{i\leq k:\ \mathrm{sign}(X^{(1)}_{\rho(i)}) \neq \mathrm{sign}(X^{(2)}_{\rho(i)})\}|/k$ for each $k \in \{1,\ldots n\}$ \Comment{IR-SDPs}
    \State $k^* \gets \max\{k:\ U_k \leq \beta q\}$ \Comment{Identify cutoff}
    \State $\hat S \gets \{\rho(1),\rho(2),\ldots, \rho(k^*)\}$ \Comment{Select subset of parameters}
    \State $\hat y_i \gets \mathrm{sign}(X^{(1)}_i)$ for each $i \in \hat S$ \Comment{Sign estimates}
    \State \Return $\hat S, \boldsymbol{\hat y}$
\end{algorithmic}
\end{algorithm}

We propose a method, called ``Sort, Select, Estimate'' (SSE), to identify a subset of parameters whose signs can be reliably estimated and estimate their signs (\cref{alg:SSE}).  The output of SSE is a subset $\hat S$ of selected parameters and a sign estimate $\hat y_i$ for each $i \in \hat S$.    

\subsection{SSE input}
A vector comprising a single measurement of each of the parameters is considered a ``replicate.''  We assume access to exactly $R=2$ replicates.  To generalize our analysis to the case $R>2$, measurements can be averaged to produce exactly two ``aggregated measurements'' for each parameter.\footnote{See \cref{sec:analysis} for a discussion of how different values of $R$ affect method performance and \cref{subsec:sims:IIIIIIR} for an investigation of varying $R$ through simulations.}
As input, SSE takes these two independent replicates: $\boldsymbol{X}^{(1)}=(X^{(1)}_1,\ldots,X^{(1)}_n)$ and $\boldsymbol{X}^{(2)}=(X^{(2)}_1,\ldots,X^{(2)}_n)$.  The measurements in the first replicate are referred to as the ``proposal measurements'' and the measurements in the second replicate are referred to as the ``validation measurements.'' 
SSE treats the two replicates asymmetrically, but the method controls type S error regardless of which replicate is used for proposal measurements and which replicate is used for validation measurements.

As input, SSE also requires a faithfulness level $q$ that the measurements are assumed to satisfy, according to the following definition.
\begin{definition}
    A random variable $B$ is a $q$-faithful measurement for a parameter $\eta$ if $\mathbb{P}(\mathrm{sign}(B)=\mathrm{sign}(\eta)) \geq q$.  
\end{definition}
In our simulations and case studies, we assume $1/2$-faithful measurements.  This is typically a mild assumption: each sign estimator outperforms a fair coin toss. 

Finally, as input, SSE requires a target type S error proportion $\beta$.  The method is designed to control the type S error proportion in the estimates $\{\hat y_i\}_{i \in \hat S}$ below $\beta$ (cf.~\cref{sec:analysis} for formal guarantees).  For example, if $\beta =0.05$, we may expect that at least 95\% of the sign estimates produced by SSE are correct.  

\subsection{SSE inner workings}
First, SSE calculates a ranking $\rho$ on the proposal measurements in order of descending magnitude (cf.~\cref{alg:SSE}, line 1).  If different treatment effect parameters have incomparable units, an alternative ranking strategy may be more suitable; we discuss the role of magnitudes in the SSE at the end of \cref{sec:analysis}.    Next, for various subsets of parameters, SSE calculates an inter-replicate sign disagreement proportion, which we define next.
\begin{definition}
    The sign disagreement proportion (SDP) between two vectors $\boldsymbol{b} \in \mathbb{R}^n$ and $\boldsymbol{c} \in \mathbb{R}^n$ is $\mathrm{SDP}(\boldsymbol{b},\boldsymbol{c})=|\{i:\ \mathrm{sign}(b_i)\neq \mathrm{sign}(c_i)|/n$.  
\end{definition}

\begin{definition}
    The inter-replicate sign disagreement proportion (IR-SDP) for the subset $\hat S\subset \{1,\ldots n\}$ is the sign disagreement proportion between $\{X^{(1)}_i\}_{i \in \hat S}$ and $\{X^{(2)}_i\}_{i \in \hat S}$, that is,
    $
    |\{i \in \hat S:\ \mathrm{sign}(X^{(1)}_i) \neq \mathrm{sign}(X^{(2)}_i)| / |\hat S|.
    $
\end{definition}

SSE computes the IR-SDP for each of the following subsets of indices (cf.~line 2):
$
\{\},\,\, \{\rho(1)\},\,\, \{\rho(1),\rho(2)\},\,\, \ldots,\,\, \{\rho(1)\ldots,\rho(n)\}.
$
\cref{tab:sdrexample} provides a worked example of this sequence of IR-SDP computations.  Then, SSE identifies the largest subset whose IR-SDP is less than $\beta q$ (cf.~lines 3 and 4).  Finally, SSE uses the proposal measurements to estimate the sign for each parameter in the subset (cf.~line 5).

\begin{table}
    \centering
    \setlength{\tabcolsep}{2pt}
    {\fontsize{9}{7}\selectfont  \begin{tabular}{l|cccccccccccccccccc}
\toprule
Parameter & $\theta_{\rho(1)}$ & $\theta_{\rho(2)}$ & $\theta_{\rho(3)}$ & $\theta_{\rho(4)}$ & $\theta_{\rho(5)}$ & $\theta_{\rho(6)}$ & $\theta_{\rho(7)}$ & $\theta_{\rho(8)}$ & $\theta_{\rho(9)}$ & $\theta_{\rho(10)}$ & $\theta_{\rho(11)}$ & $\theta_{\rho(12)}$ & $\theta_{\rho(13)}$ & $\theta_{\rho(14)}$ & $\theta_{\rho(15)}$ & $\theta_{\rho(16)}$ & $\theta_{\rho(17)}$ \\
\midrule
$X^{(1)}$ & -8.5 & -7.7 & -7.1 & 7.1 & -6.8 & 6.4 & 4.8 & -4.7 & -4.4 & 4.0 & 3.9 & \textcolor{red}{3.2} & -3.1 & 2.4 & -1.3 & \textcolor{red}{1.3} & 0.8 \\
$X^{(2)}$ & -5.8 & -5.7 & -9.1 & 1.5 & -3.1 & 8.9 & 1.9 & -7.5 & -5.8 & 5.2 & 1.3 & \textcolor{red}{-3.3} & -6.3 & 6.2 & -4.4 & \textcolor{red}{-2.6} & 6.9 \\
SDP & 0.0\% & 0.0\% & 0.0\% & 0.0\% & 0.0\% & 0.0\% & 0.0\% & 0.0\% & 0.0\% & 0.0\% & 0.0\% & 8.3\% & 7.7\% & 7.1\% & 6.7\% & 12.5\% & 11.8\% \\
\bottomrule
\end{tabular}}
    \setlength{\tabcolsep}{6pt}
    \caption{A worked example of the SSE method.
    Sort the parameters using the magnitudes of the proposal measurements, $\boldsymbol{X}^{(1)}=(\boldsymbol{X}^{(1)}_1,\ldots,\boldsymbol{X}^{(1)}_n)$.  For each $i$, determine whether the sign of the proposal measurement $X^{(1)}_i$ is the same as the sign of the validation measurement $X^{(2)}_i$; the red color indicates sign disagreements.  For each $k\in \{1,\ldots,n\}$, compute the inter-replicate sign disagreement proportion (IR-SDP) for the first $k$ parameters.  SSE selects parameters using these IR-SDPs and a user-specified target error level.  
    \label{tab:sdrexample}}
\end{table}

\section{Analysis of SSE's error control}
\label{sec:analysis}

We now analyze the type S error control of the SSE method.  The output of the SSE method is a subset of parameters, $\hat S$, together with a sign estimate $\hat y_i$ for each $i \in \hat S$.  The type S error proportion of this output is the sign disagreement proportion between $\{\hat y_i\}_{i \in \hat S}$ and the true parameters $\{\theta_i\}_{i \in \hat S}$.  

To analyze the SSE, we first demonstrate a linear inequality about sign disagreement proportions with faithful measurements: the sign disagreement proportion between a vector $\boldsymbol{c}$ and the parameters $\boldsymbol{\eta}$ can be controlled in terms of the expected sign disagreement proportion between $\boldsymbol{c}$ and a collection of $q$-faithful measurements $\boldsymbol{B}$.

\begin{lemma} \label{thm:lowertardis}
    For $i\in \{1,\ldots,k\}$, let $B_i$ be a $q$-faithful measurement for parameter $\theta_i$.   Then, for any deterministic vector $\boldsymbol{c}=(c_1,\ldots,c_k)$, 
    $
    \mathrm{SDP}(\boldsymbol{c},\boldsymbol{\theta}) \leq \mathbb{E}[\mathrm{SDP}(\boldsymbol{c},\boldsymbol{B})]/q.
    $
\end{lemma}

We next bound the type S error proportion of the output of the SSE method, by applying \cref{thm:lowertardis} with a subset of proposal measurements supplying the values for $\boldsymbol{c}$ and a subset of validation measurements supplying the values for $\boldsymbol{B}$.  To use this lemma, our analysis treats the proposal measurements as non-random (or, equivalently, conditioned upon and independent of the validation measurements).  In particular, for any fixed proposal measurements, we can use \cref{thm:lowertardis} to show that the expected value of each IR-SDP (which quantifies disagreement between two replicates in some subset) can be used as a proxy for a corresponding type S error proportion (which quantifies disagreement between the measurements and the truth in some subset).  \cref{thm:asymtopia}, below, uses this proxy to show that the SSE method asymptotically maintains the type S error proportion below its nominal target in the limit of large $n$, under conditions ensuring that the number of discoveries grows with $n$ and the dependencies among the validation measurements are well-controlled.  Note that, although the theorem considers the proposal measurements to be non-random, type S error control of the SSE method with random proposal measurements follows immediately from the dominated convergence theorem.  

\begin{theorem} \label{thm:asymtopia}
Let $\boldsymbol{W}^{(1)}=\left(W_{1}^{(1)}, W_{2}^{(1)},\ldots\right)$ denote any fixed (non-random) infinite sequence of measurements and let $\boldsymbol{W}^{(2)}=\left(W_{1}^{(2)},W_{2}^{(2)},\ldots\right)$ denote an infinite sequence of random measurements.  For a fixed faithfulness $q\in(0,1)$ and each $i\geq1$, assume that $W_{i}^{(2)}$ is a $q$-faithful measurement for parameter $\theta_{i}$. 
Fix a target type S error level $\beta \in (0, 1)$.   

For each $i \geq 1$, let $D_{i}=\mathbb{I}\left(\mathrm{sign}(W_i^{(1)})\neq \mathrm{sign}(W_i^{(2)})\right)$ denote an indicator for a sign disagreement regarding the parameter $i$. For each $n\geq 1$, let $\rho_{n}$ denote the permutation of $(1,\ldots, n)$ that indicates the descending rank statistics of $\left(\left|W_{1}^{(1)}\right|,\ldots,\left|W_{n}^{(1)}\right|\right)$.
For each $n\geq 1$ and $k\in\{1,\ldots,n\}$, let $U_{n,k}=\frac{1}{k}\sum_{i=1}^{k}D_{\rho_{n}(i)}$ denote an IR-SDP for the subset $\{\rho_n(1),\ldots,\rho_n(k)\}$.  
Fix any $\delta>0$. For each $n\geq 1$, let $N_{n}=\max\left\{ k\leq n:\ \mathbb{E}\left[U_{n,k}\right]\leq\beta q-\delta\right\}$, so that the expected IR-SDP for the subset $\{\rho_n(1),\ldots,\rho_n(N_n)\}$ lies below $\beta q-\delta$. Assume that $\lim_{n\rightarrow\infty}N_{n}=\infty$.  Also assume the following dependency condition: there exists a constant $c$ such that for every $T \subset \mathbb{N}$, $\sum_{\substack{i\neq j\\ i,j \in T}
}\mathrm{cov}\left(D_{i},D_{j}\right)\leq c|T|$.

For each $n\geq 1$, let $V_n$ denote the type S error proportion of the output of the SSE method (\cref{alg:SSE}) applied with arguments $\left(W_{1}^{(1)},\ldots,W_{n}^{(1)}\right)$, $\left(W_{1}^{(2)},\ldots,W_{n}^{(2)}\right)$, $\beta$, and $q$. 
Then, $$\lim_{n\rightarrow\infty}\mathbb{P}\left(V_{n}>\beta+\delta/q\right)=0.$$
\end{theorem}

\begin{remark}
    The dependency condition in \cref{thm:asymtopia} holds naturally in high-throughput experiments, which are often divisible into independent sub-experiments, each measuring a smaller set of parameters.  The covariances $\{\mathrm{cov}\left(D_{i},D_{j}\right)\}_{i,j}$ then admit a block-diagonal structure, and the dependency condition is satisfied as long as the size of the blocks is always bounded by some constant $M$ even as the number of parameters grows.
\end{remark}

\begin{remark}
    To apply the SSE if $R\geq 2$, measurements can be averaged to produce exactly two ``aggregated measurements'' for each parameter.  In most cases we expect that this averaging will produce aggregated measurements that tend to exhibit higher IR-SDP values (i.e., increased sign agreement).  Increased IR-SDP values will generally lead to more discoveries.  Regardless, the type S error guarantee will be unaffected.  
\end{remark}

\begin{remark}
    The proof for \cref{thm:asymtopia} is unchanged regardless of how the SSE computes the ordering $\rho$ from the proposal measurements.  However, the ordering is crucial for the power of the SSE: the number of discoveries will be largest if inter-replicate sign disagreements tend to lie at the end of the ordering.  In this article, we use the magnitudes of the proposal measurements to facilitate this property.  Specifically, we apply the heuristic that larger magnitudes $|X_i^{(1)}|$ indicate a higher likelihood that the signs of $X_i^{(1)}$ and $X_i^{(2)}$ agree.  Under this heuristic, choosing $\rho$ by magnitude will typically place sign disagreements in the ordering towards the end.  However, alternative orderings may be preferred if different measurements use incomparable units or if other heuristics are available.  The user may supply any ordering that is independent of the validation measurements.
\end{remark}


\section{Simulations}
\label{sec:simulations}

We use simulations to probe the strengths and weaknesses of the proposed SSE method (\cref{alg:SSE}).
We first evaluate performance with a variety of different noise distributions.  We then compare SSE with methods based on the existing literature (\cref{subsec:sims:badmodels}).  
Finally, we use simulations to evaluate various ways of applying our techniques to experiments that produce more than two measurements of each parameter (\cref{subsec:sims:IIIIIIR}).

\subsection{Performance with different noise distributions}
\label{subsec:sims:IIIIII}

\paragraph{Simulation.}  The parameters of interest comprise a matrix $\boldsymbol{\theta} \in \mathbb{R}^{P \times G}$ indicating the effects of each perturbation $p \in \{1,\ldots, P\}$ on each gene $g \in \{1,\ldots,G\}$.  The true effects are simulated at random.  The effect magnitudes are simulated to be non-uniform over perturbations, such that parameters in the upper rows of $\boldsymbol{\theta}$ involve smaller effects and parameters exhibit progressively larger effects as the row index increases (cf.~line 2 of \cref{alg:SSE}).\footnote{Note that this ordering of the perturbations will not be used by any of the methods; each method will re-order parameters according to the magnitudes of $X^{(1)}$ alone.}  The noise distribution has two components.  The first noise component is a low-rank Gaussian component, inducing dependency across estimates concerning the same perturbation.  This component is intended to simulate low-rank correlation structures, which are widely believed to describe gene regulatory programs \citep{ye2013low}. An integer constant $K$ controls the rank of the low-rank component and a constant $\sigma_1^2$ controls the magnitude of the low-rank noise.  The second noise component is a Student's t distribution that is independent across all estimates. A constant $\sigma_2$ controls the magnitude of this independent noise.  The final estimates for each replicate are given by adding the true parameter value to the noise.  The exact simulation procedure can be found in \cref{alg:simIIIIII} of \cref{apdx:simprocess}.

In this section, we fix $P=1000$, $G=200$, $K=10$, and $R=2$, and consider simulations with various choices for $\sigma^2=\sigma_1^2+\sigma_2^2$, $k=\sigma_1^2/\sigma_2^2$, and the degrees of freedom in the Student's t distribution.  We apply the SSE method targeting a 10\% type S error proportion.

\paragraph{Results.} The number of discoveries across different simulation conditions and different simulators can be found in \cref{tab:eIIIIII}.  Fewer degrees of freedom lead to fewer discoveries.  Higher noise leads to fewer discoveries.   Greater low-rank noise and additional dependencies lead to fewer discoveries.   The method controls type S error to the nominal level in all cases. We targeted a 10\% type S error proportion, but the maximum false discovery sign proportion observed in any simulation was only 3\%.  Conservative error control may be the price of using a model-free method.  

\begin{table}
    \centering
    {
    \small
    \begin{tabular}{ll|rrrrrrrrr}
\toprule
 & $\sigma^2$ & \multicolumn{3}{r}{0.1} & \multicolumn{3}{r}{0.2} & \multicolumn{3}{r}{0.3} \\
 & $k$ & 0.0 & 1.0 & 2.0 & 0.0 & 1.0 & 2.0 & 0.0 & 1.0 & 2.0 \\
\midrule
\multirow[t]{3}{*}{} & $d=3$ & 156.2 & 147.5 & 146.0 & 101.1 & 90.8 & 89.9 & 51.1 & 51.2 & 49.6 \\
 & $d=5$ & 150.1 & 146.6 & 145.0 & 92.7 & 90.4 & 89.3 & 48.2 & 50.8 & 50.4 \\
 & $d=7$ & 148.3 & 146.2 & 144.7 & 90.0 & 89.6 & 88.7 & 48.0 & 49.9 & 49.8 \\
\cline{1-11}
\end{tabular}
    }
    \caption{Number of discoveries (in thousands).    We consider simulations including different degrees of freedom for the noise distribution ($d$), different levels of noise ($\sigma$), and different ratios of low-rank noise to independent noise ($k$).}
    \label{tab:eIIIIII}
\end{table}

\subsection{Comparison with extant methods}
\label{subsec:sims:badmodels}
We next conduct simulations to compare SSE with existing methods.  Ideally, we would compare with other methods from the literature that also provably control type S error in our setting.  However, there are no such methods.  We therefore compare with two competitors from the literature that might control type S error in our setting.

\paragraph{Competitor 1.}

The first competitor, which we refer to as Misspec-BH,
combines a simple misspecified parametric model with the method of \citet{benjamini_false_2005} to control the expected type S error proportion. Consider a model in which $\theta_i$ is distributed according to $\mathcal{N}(\theta_i,\tilde \tau^2)$.  This model has been considered by \citet{zhao2020intrigue} for similar data. This model matches the true data-generating process of our simulation, except that the model assumes that the variance of every measurement is the same (whereas different measurements have different variances in our simulations).  Remedying this misspecification would be difficult: it would be challenging to estimate the variance for all $n$ parameters because there are only two independent observations for each.   We estimate $\tau^2$ and then apply the method of \citet{benjamini_false_2005} to $p$-values based on this misspecified model, with a target expected type S error proportion of 10\%.  This method ensures that the expected type S error proportion is below the target level as long as (i) the distribution of $X_i$ is correctly modeled for all $i$ such that $\theta_i=0$  and (ii) the distribution of each $X_i$ increases stochastically as a function of $\theta_i$.  Note that these assumptions are not met except when the simulation constant $k$ takes the value $k=1$, because the misspecified parametric model does not correctly model the distribution of $X_i$ under the null hypothesis that $\theta_i=0$.  We compare with this method, even though its assumptions are not met, because (i) the approach of \citet{benjamini_false_2005} is widely practiced and (ii) it may perform well even though its assumptions are not met.  

\paragraph{Competitor 2.}
The second competitor, which we refer to as Quad-IDR, is based on the Irreproducible Discovery Rate  ({\IDR}) procedure developed by \citet{li2011measuring}.   The original IDR procedure uses two scalar confidence scores for each parameter, obtained from two independent measurements.  It is assumed that the copula of confidence scores can be modeled as a two-component Gaussian mixture model.   The parameters of this mixture model are fit.  For each parameter, the posterior probability that the parameter arose from the mixing component corresponding to lower confidence scores is termed the local irreproducible discovery rate.  Practitioners are encouraged to focus on parameters with smaller local irreproducible discovery rates.  The IDR procedure was not originally designed to handle the estimation of parameter signs.  However, it is one of the few existing methods that can accommodate invalid $p$-values.

We invent a new procedure, Quad-IDR, that uses the original IDR procedure to estimate signs.  Quad-IDR applies the original IDR procedure twice, based on the ``quadrants'' of each pair of measurements: it applies the IDR once to the parameters where both measurements were positive and once to the parameters where both measurements were negative. It ignores parameters for which the measurements had different signs.

\paragraph{Simulations.}
In each simulation, we drew $n=50,000$ parameter values, $\theta_1\ldots \theta_n$, from a standard normal distribution.  We then simulated two replicates of a study to use in estimating the parameters.  We assumed that the measurement of each replicate for each $\theta_i$ was distributed according to $\mathcal{N}(\theta_i,\tau_i^2)$.  We constructed the variances $\tau_i^2$ in terms of two constants: $\sigma$ and $k$.  For 90\% of the parameters, we took the variance to be $\sigma^2$; for 10\% of the parameters, we took it to be $k \sigma^2$.  With this parameterization, by selecting $k \ne 1$, we can simulate studies that are better suited for estimating some parameters than others.  We varied $\sigma$ between $0.1$ and $1.0$ and $k$ between $1$ and $10$.  In each simulation, we attempted to infer the signs of $\boldsymbol{\theta}$. 

\begin{figure}
\begin{center}
\includegraphics[width=\figwidth]{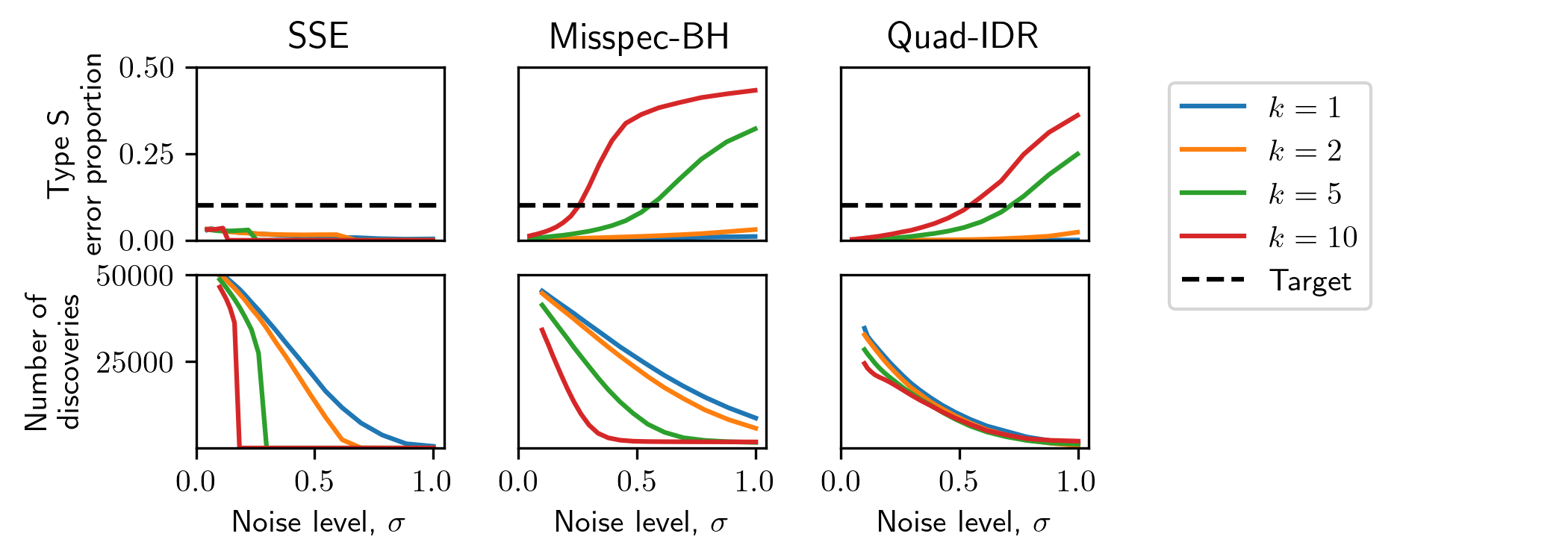}
  \caption{
  \captionheading{Comparing SSE with competitors for error control in simulated data.}  When $k$ and $\sigma^2$ are large, it is challenging to accurately estimate the signs of the parameters.  In this regime, the SSE made no discoveries.  The misspecified Benjamini-Hochberg method and quadrant IDR method made some discoveries, but a high proportion of these were false discoveries.
  }\label{fig:sims}
\end{center}
\end{figure}

\paragraph{Results.}
Figure \ref{fig:sims} compares SSE, Misspec-BH, and Quad-IDR.  All made many true discoveries when the noise level $\sigma^2$ was not too high.  All methods also controlled error at the target level when $k=1$; in this case, at least the second method's model was specified correctly.  However, when $k$ and $\sigma^2$ were large, Misspec-BH had sign error proportions as high as 43\% and Quad-IDR had sign error proportions as high as 36\%.  These methods failed to control type S error proportions at their target levels.  SSE always controlled type S error, keeping it below its target level.

\subsection{Using more than two independent replicates}

Some high-throughput experimental designs include more than two replicates.  When the number of replicates $R$ is greater than two, our SSE method cannot be applied directly.  Instead, we suggest averaging $R_p$ replicates to form the proposal measurements and $R-R_p$ replicates to form the validation measurements.  

The number of discoveries made by each method may depend on the choice of $R_p$. If $R_p$ is too small, the proposal measurements will suffer in accuracy. If $R_p$ is too large, the validation measurements will suffer in accuracy.  Here, we use simulations to assess what values of $R_p$ may be suitable in practice.  

Note that the number of discoveries may depend on \emph{which} replicates are used as well as \emph{how many} replicates are used in forming the proposal and validation measurements.  In this article, because we assume that the replicates are all produced by the same experimental protocol, this choice can be made arbitrarily.

\paragraph{Simulations.}
We constructed simulated datasets using the same simulation process from \cref{subsec:sims:IIIIII}, with $P=1000$, $G=200$, $K=10$, and $\sigma_1=0$.  We explored different values of $\sigma_2$.  We considered the case $R=5$ (in which case $R_p \in \{1,2,3,4\}$) and the case $R=3$ (in which case $R_p \in \{1,2\})$).  For each simulation, we merged replicates by averaging measurements and applied SSE with a target type S error control of 10\%.

\paragraph{Results.}
The type S error proportion was always below 3.4\%, well below the nominal level. \cref{tab:eIIIIII_rep} indicates that higher noise levels lead to fewer discoveries and the best choice for $R_p$ is given by $R_p = \lceil R/2 \rceil$.  In all cases, setting $R_p = \lceil R/2 \rceil$ leads to more discoveries than $R_p = \lfloor R/2 \rfloor$.  For example, with $\sigma_2=1$ and $R=5$, setting $R_p = \lceil R/2 \rceil$ yields 30 times more discoveries. We conjecture that it is preferable to use slightly more replicates to form the proposal measurements, as these measurements are used for two purposes (for sorting parameters and for estimating signs).  

\label{subsec:sims:IIIIIIR}
\begin{table}
    \centering
    {
    \small
    \hfill{}
    \begin{tabular}{l|rrrrr}
    \toprule
    $\sigma_2$ & 0.2 & 0.4 & 0.6 & 0.8 & 1.0 \\
    \midrule
    $R_p=1$ & 109.1 & 38.3 & 4.1 & 0.0 & 0.0 \\
    $R_p=2$ & 127.5 & 64.4 & 24.8 & 6.9 & 0.1 \\
    $R_p=3$ & 129.5 & 68.2 & 30.4 & 11.4 & 3.1 \\
    $R_p=4$ & 118.6 & 55.8 & 22.4 & 6.9 & 1.5 \\
    \bottomrule
    \end{tabular}
    \hfill{}
    \begin{tabular}{l|rrrrr}
    \toprule
    $\sigma_2$ & 0.2 & 0.4 & 0.6 & 0.8 & 1.0 \\
    \midrule
    $R_p=1$ & 102.0 & 32.7 & 1.0 & 0.0 & 0.0 \\
    $R_2=2$ & 106.1 & 41.8 & 11.7 & 0.8 & 0.1 \\
    \bottomrule
    \end{tabular}
    \hfill{}
    }
    \caption{Number of discoveries (in thousands) across different ways of using the replicates.  On the left, we consider a setting with 5 replicates, in which case $R_p \in \{1,2,3,4\}$. On the right, we consider a setting with three replicates, in which case $R_p \in \{1,2\}$.  In both cases, we consider simulations with different levels of noise ($\sigma_2$).}
    \label{tab:eIIIIII_rep}
\end{table}

\ignoreme{
 _____  ______          _      _____       _______
|  __ \|  ____|   /\   | |    |  __ \   /\|__   __|/\
| |__) | |__     /  \  | |    | |  | | /  \  | |  /  \
|  _  /|  __|   / /\ \ | |    | |  | |/ /\ \ | | / /\ \
| | \ \| |____ / ____ \| |____| |__| / ____ \| |/ ____ \
|_|  \_\______/_/    \_\______|_____/_/    \_\_/_/    \_\

}

\section{Case studies}
\label{sec:casestudies}

To probe the strengths and weaknesses of our model-free methods for error assessment and control, we conducted a series of case studies using data from the L1000 platform.  \citet{subramanian2017next} used this platform to investigate the upregulation and downregulation of various genes in the presence of various small-molecule perturbations.  The platform applies different perturbations to many wells of cells.  A fluorescent marking system measures the total abundance of 978 distinct types of RNA in each well. The raw fluorescent measurements for a given well cannot be interpreted directly, but must be processed to obtain what is referred to as a ``$z$-score'' for any given perturbation-gene combination, indicating whether the gene is upregulated or downregulated by the perturbation.  Positive $z$-scores indicated upregulation and negative $z$-scores indicated downregulation.  It is unclear whether these $z$-scores were associated with valid hypothesis tests. \citet{qiu2020bayesian} developed a new method for processing raw fluorescent measurements to improve the signal-to-noise ratio in the $z$-scores.  We refer to the original method as {\LonekA} and the new method as {\LonekB}.  We assume that both methods produce $1/2$-faithful measurements, i.e., the sign of each $z$-score agrees with the true direction of differential expression at least 50\% of the time.\footnote{``Differential expression'' is not precisely defined by \citet{subramanian2017next}, but we assume it to mean the expected change in total RNA abundance for a given gene among wells that have undergone a given perturbation.} 

\citet{subramanian2017next} used the L1000 platform to study perturbations on several different populations of cells, including the A375 cell line.  The A375 cell line is a population of cells descending from a human melanoma that has been stabilized to persist indefinitely without change in lab conditions.  For this cell line, \citet{subramanian2017next} investigated the effects of 1,482 different perturbations.  Some of these molecules are known to impair a variety of cancerous cells.  Some are not completely understood.  Several wells were used to study each perturbation.  Treating each well as independent, this data matches the setting for the SSE method: a small number of independent measurements of each parameter, with many parameters measured in total.

We focus on the estimand $\boldsymbol{\theta} \in \mathbb{R}^{1,482 \times 978}$ considered by \citet{subramanian2017next} for the A375 cell line.  We compare the independent $z$-score measurements produced by {\LonekB} and the independent measurements produced by {\LonekA} (\cref{subsec:casestudies:preproc}).  We use the theoretical results of \cref{sec:method} to support the claim of \citet{qiu2020bayesian} that the {\LonekB} measurements are more accurate.  We then used the SSE method to estimate $\mathrm{sign}(\boldsymbol{\theta})$, using {\LonekB} data (\cref{subsec:casestudies:L1000}).  We compare SSE with the Benjamini-Hochberg (BH) method of \citet{benjamini_false_2005}, applied under the assumption that each ``$z$-score'' is in fact a unit-variance measurement.  The assumption does not appear to be met in this case, highlighting the need for model-free methods in this context.

\subsection{Comparison of {\LonekA} and {\LonekB}}
\label{subsec:casestudies:preproc}

\citet{qiu2020bayesian} claimed that the {\LonekB} method for processing fluorescence was more accurate.  They based this claim on the observation that it yielded more consistent answers across replicates than {\LonekA} did.  However, \citet{qiu2020bayesian} did not have a rigorous way to connect their observation about estimator replicability to a notion of accuracy.  \Cref{thm:lowertardis} provides the necessary link. 

We first applied SSE to both {\LonekA} and {\LonekB} to discover sign estimates, targeting a type S error proportion of 10\%.  Based on our findings from \cref{subsec:sims:IIIIIIR}, we used two replicates to generate proposal measurements and one replicate for the validation measurements.  We merged replicates by averaging the $z$-scores.  

\begin{figure}
\begin{center}
\includegraphics[width=\figwidth]{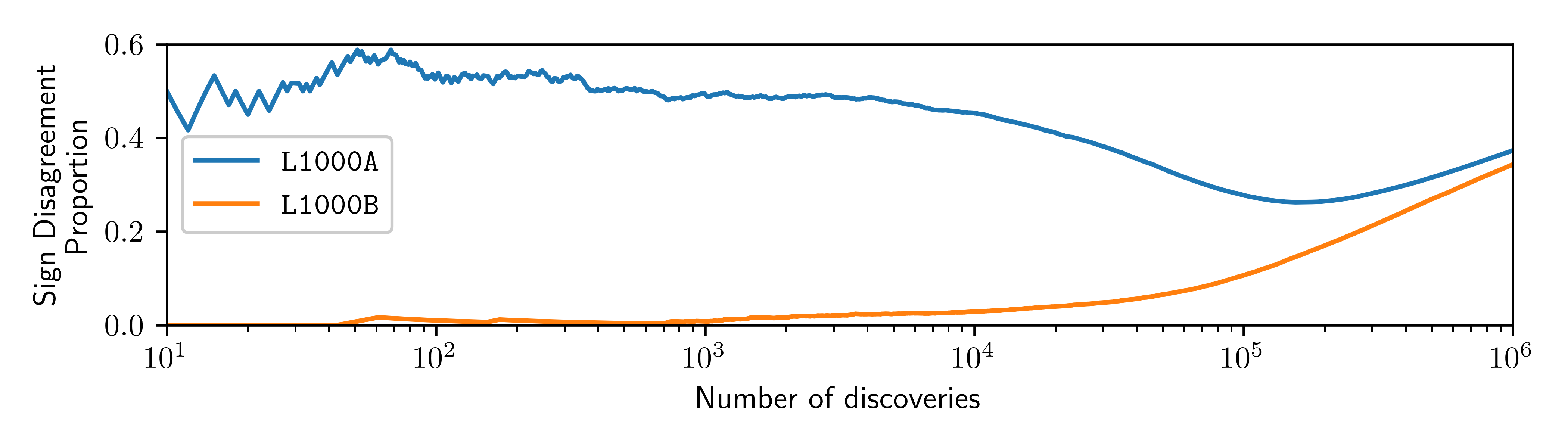}
  \caption{
  \captionheading{The SDP suggests lower type S error proportions in the estimates of a new protocol.}  For each $k$, we compute the SDP for {\LonekA} and {\LonekB} estimates of the parameters $\{\theta_j\}_{j<k}$.
  }\label{fig:L1000preproc}
\end{center}
\end{figure}

Of the $1,482\times 987=1,449,396$ parameters considered, SSE yielded 2 discoveries when applied to {\LonekA} and 32,964 discoveries when applied to {\LonekB}.  This stark difference is due to the fact that, as noted in \citet{qiu2020bayesian}, the measurements from {\LonekB} were much more consistent across replicates.  For each $k$, \Cref{fig:L1000preproc} shows the SDP between the replicates for the parameters $\{\theta_j\}_{j \leq k}$.  For example, in {\LonekB} with $k=100,000$ parameters we observe an {\SDP} of 5\%, suggesting a type S error proportion below 10\%.   On the other hand, in {\LonekA} measurements, the SDP is greater than 25\% for all choices of $k$.  Moreover, the largest {\SDP}s in the {\LonekA} data appear among subsets of parameters with the largest magnitudes of proposal measurements; such subsets are shown toward the left side of \Cref{fig:L1000preproc}.  We might expect measurements with high magnitudes to correspond to stronger effects with fewer errors, so this result is somewhat surprising.  This evidence supports the conjecture from \citet{qiu2020bayesian} that a small proportion of the fluorescence measurements have extreme measurement errors, and thus that the original heuristic algorithm overestimates effect sizes and underestimates error levels in these anomalous readings.
 
Although the original L1000 data was published in 2017, it took four years for serious concerns about its accuracy to rise to the level of a publication \citep{lim2021evaluation}.  If the original experimentalists had calculated the {\SDP}, some of these problems could have been identified immediately.  
Developing and implementing protocol-specific strategies to evaluate error is difficult.  Simple best practices for assessing error that can be applied across experimental modalities would alleviate this burden.

\subsection{Estimating signs from L1000}

\label{subsec:casestudies:L1000}

We now seek to estimate signs from L1000 data: to identify a subset of parameters where the sign can be reliably inferred and a sign estimate for each parameter in the subset.  We first attempt this estimation using traditional techniques and then turn to the SSE method.  

We begin by assuming the $z$-scores produced by {\LonekA} and {\LonekB} are distributed according to the standard normal distribution under the null hypothesis of no effect.  We further assume that the distribution of each $z$-scores is monotone stochastically increasing with its estimand.  If these assumptions are met, the procedure of \citet{benjamini_false_2005} will control the expected type S error proportion.  We apply that procedure, targeting an expected type S error proportion of 10\%, and obtain 228,005 discoveries from {\LonekA} and 23,086 discoveries from {\LonekB} data.  From this, we might surmise that the measurements from {\LonekA} were substantially more informative.  However, this conclusion would only be warranted if the $z$-scores from both {\LonekA} and {\LonekB} were indeed governed by the standard normal distribution under the null hypothesis of no effect.  Given the results from \cref{subsec:casestudies:preproc} suggesting that {\LonekB} may in fact contain more accurate measurements, we conjecture the $z$-scores may not be governed by the standard normal distribution under the null hypotheses.  Indeed, although the measurements of {\LonekA} and {\LonekB} are referred to as $z$-scores, we suppose it may be quite difficult in practice to produce valid $z$-scores from the fluorescence measurements of the L1000 system.  Doing so would require correctly modeling batch effects, which can arise from a variety of nuanced unmeasured aspects of the experimental process (e.g., lab technician proficiency, humidity).

We now consider the possibility that the $z$-scores produced by {\LonekA} and {\LonekB} are not in fact governed by the standard normal distribution under the null hypotheses.  To perform estimation in this setting, we use the SSE method, which is model-free.  We average two replicates to produce proposal measurements and one to produce validation measurements.  We focus on {\LonekB} data as \cref{subsec:casestudies:preproc} suggested it was more reliable.  

Using SSE, targeting a 10\% type S error proportion, we obtained 38,283 discoveries in {\LonekB}.  This represents 2.6\% of the $978 \times 1482$ parameters investigated, and considerably more than the 23,086 parameters discovered by BH.  The discoveries were not spread evenly across the 978 perturbations.  For example, 365 perturbations were associated with zero discoveries.  In contrast, 115 perturbations were associated with 100 or more discoveries.  This suggests that some perturbations have minimal effect, while others affect many genes.  \cref{apdx:L1000discbypert} shows the distribution of discoveries across different perturbations.  

\ignoreme{
 ____ ___ ____   ____ _   _ ____ ____ ___ ___  _   _
|  _ \_ _/ ___| / ___| | | / ___/ ___|_ _/ _ \| \ | |
| | | | |\___ \| |   | | | \___ \___ \| | | | |  \| |
| |_| | | ___) | |___| |_| |___) |__) | | |_| | |\  |
|____/___|____/ \____|\___/|____/____/___\___/|_| \_|

}
 
\section{Discussion}
\label{sec:discussion}

This article developed new methods for assessing and controlling type S error under an assumption of $q$-faithful measurements; this assumption is typically mild for modest values of $q$ (e.g., $q=1/2$).  However, these methods come with three limitations.

First, we caution that even the modest $1/2$-faithful assumption could fail if the measurement bias or skew substantially exceeds the magnitude of the true effect.  It is not possible to directly assess whether the assumption holds using the measurements $\boldsymbol{X}$ alone.  To determine an appropriate constant $q$ so that the measurements can be assumed to be $q$-faithful, it may be helpful to conduct low-throughput high-accuracy trials to obtain ground truth about some of the parameters of interest.

Second, our theoretical results about SSE are asymptotic in nature.  If the cutoff $k^*$ identified by the SSE method is small, the user may not be in the asymptotic regime: type S error control may not be achieved.  We leave finite-sample guarantees for future work.


Third, the model-free nature of the proposed method limits the method's power.  In simulations, we considered cases where valid $p$-values were available.  In these cases, applying traditional methods with those $p$-values leads to more discoveries relative to the SSE method that we propose.  To improve power, one could incorporate prior knowledge using Bayesian techniques to obtain more accurate estimates for the proposal measurements (SSE controls error regardless of how the proposal measurements are produced, as long as they remain independent of the validation measurements).  This suggests a promising direction for improving the power of model-free methods for cell-perturbation data.  

\if1\blind{

\textbf{Software.}  Notebooks for reproducing all experiments in this article are included with the online version of this article.

\textbf{Acknowledgements.}  We thank Robert Barton, Meena Subramaniam, Maxime Dhainaut, and Drausin Wulsin for introducing us the scientific problems that motivated this research.

\textbf{Funding.}  This research was funded by Immunai Inc.
}\fi

\textbf{Supplementary material.}  Appendices (including one that contains the proofs) are included in the supplementary material.


\bibliographystyle{plainnat}
\bibliography{refs}

\newpage


\begin{center}
    \huge Supplementary Material for ``Model-free Sign Estimation for High-Throughput Screenings%
''
\end{center}

\appendix

\section{Proofs}

\begin{lemma*}[Restatement of \cref{thm:lowertardis}] If $\boldsymbol{Y}\in \{-1,1\}^n$ are faithful validation signs for $\boldsymbol{\truprop}$, then the type S error proportion is bounded by $V(\boldsymbol{\hat y},\boldsymbol{\truprop}) \leq \mathrm{{\SDR}}(\boldsymbol{\hat y})/q$.
\end{lemma*}

\begin{proof}

Observe that
\begin{align*}
\mathrm{\SDR} & =\frac{1}{n}\left(\sum_{i:\ \hat{y}_{i}=\mathrm{sign}(\theta_{i})}{\Prsymbol}\left(Y_{i}\neq\hat{y}_{i}\right)+\sum_{i:\ \hat{y}_{i}\neq\mathrm{sign}(\theta_{i})}{\Prsymbol}\left(Y_{i}\neq\hat{y}_{i}\right)\right)\\
 & =\frac{1}{n}\left(\sum_{i:\ \hat{y}_{i}=\mathrm{sign}(\theta_{i})}{\Prsymbol}\left(Y_{i}\neq\mathrm{sign}\left(\theta_{i}\right)\right)+\sum_{i:\ \hat{y}_{i}\neq\mathrm{sign}(\theta_{i})}{\Prsymbol}\left(Y_{i}=\mathrm{sign}\left(\theta_{i}\right)\right)\right)\\
 & \geq\frac{1}{n}\left(0+\sum_{i:\ \hat{y}_{i}\neq\mathrm{sign}(\theta_{i})}q\right)\\
 & =V(\hat{\boldsymbol{y}},\boldsymbol{\theta})q.
\end{align*}

\end{proof}

\begin{remark*}
    As seen in the proof above, the bound for \cref{thm:lowertardis}
    will be conservative when $\sum_{i:\ \hat{y}_{i}=\mathrm{sign}(\theta_{i})}{\Prsymbol}\left(Y_{i}\neq\mathrm{sign}\left(\theta_{i}\right)\right)$ is large.
\end{remark*}

\begin{theorem*}[Restatement of \cref{thm:asymtopia}] 
Let $\boldsymbol{W}^{(1)}=\left(W_{1}^{(1)}, W_{2}^{(1)},\ldots\right)$ denote a fixed infinite sequence and let $\boldsymbol{W}^{(2)}=\left(W_{1}^{(2)},W_{2}^{(2)},\ldots\right)$ denote an infinite sequence of random variables.  For a fixed value $q\in(0,1)$ and each $i\geq1$, assume that $W_{i}^{(2)}$ is a $q$-faithful measurement for parameter $\theta_{i}$. 
Fix a target type S error level $\beta \in (0, 1)$.   

Now, for each $i \geq 1$, let $D_{i}=\mathbb{I}\left(\mathrm{sign}(W_i^{(1)})\neq \mathrm{sign}(W_i^{(2)})\right)$. For each $n\geq 1$, let $\rho_{n}$ denote the permutation of $(1,\ldots, n)$ that indicates the rank statistics of $\left(-\left|W_{1}^{(1)}\right|,\ldots,-\left|W_{n}^{(1)}\right|\right)$.
For each $n\geq 1$ and $k\in\{1,\ldots,n\}$, let $U_{n,k}=\frac{1}{k}\sum_{i=1}^{k}D_{\rho_{n}(i)}$.  
Fix any $\delta>0$. Let $N_{n}=\max\left\{ k\leq n:\ \mathbb{E}\left[U_{n,k}\right]\leq\beta q-\delta\right\}$. Assume that $\lim_{n\rightarrow\infty}N_{n}=\infty$ and there exists a constant $c$ such that for every $T \subset \mathbb{N}$, $\sum_{\substack{i\neq j\\ i,j \in T}
}\mathrm{cov}\left(D_{i},D_{j}\right)\leq c|T|$.

For each $n\geq 1$, let $V_n$ denote the type S error proportion of the output of the SSE method (\cref{alg:SSE}) applied with arguments $\left(W_{1}^{(1)},\ldots,W_{n}^{(1)}\right)$, $\left(W_{1}^{(2)},\ldots,W_{n}^{(2)}\right)$, $\beta$, and $q$. 
Then, $$\lim_{n\rightarrow\infty}\mathbb{P}\left(V_{n}>\beta+\delta/q\right)=0.$$
\end{theorem*}

\begin{proof}   Let $E_{n,k}$ denote the event that $\left|U_{n,k}-\mathbb{E}\left[U_{n,k}\right]\right|>\delta/2$. Let $\Sigma_{i,j}=\mathrm{cov}\left(D_{i},D_{j}\right)$.  Chebyschev's inequality, together with the boundedness condition $D_{i}\in\left\{ 0,1\right\} $, gives that 
\[
\mathbb{P}\left(E_{n,k}\right)\leq\frac{1}{k\delta^{2}}+\frac{2}{k^{2}\delta^{2}}\sum_{1\leq i<j\leq k}\Sigma_{\rho_{n}(i),\rho_{n}(j)}.
\]
Using the condition that $\sum_{\substack{i\neq j\\
i,j\in T
}
}\Sigma_{i,j}=O\left(\left|T\right|\right)$, identify $c,M$ so that so that $\sum_{\substack{i\neq j\\
i,j\in T
}
}\Sigma_{i,j}\leq c\left|T\right|$ whenever $\left|T\right|>M$. Thus
\[
\mathbb{P}\left(E_{n,k}\right)\leq\frac{1+4c}{k\delta^{2}}
\]
for all $k>M$. Now consider the event
\[
F_{n}=\bigcup_{k=\left\lfloor \sqrt{N_{n}}\right\rfloor }^{\left\lfloor \sqrt{n}\right\rfloor }E_{n,k^{2}}.
\]
For any $n$ with $\left\lfloor \sqrt{N_{n}}\right\rfloor >M$, it follows that
\begin{align*}
\mathbb{P}\left(F_{n}\right) & \leq\sum_{k=\left\lfloor \sqrt{N_{n}}\right\rfloor }^{\infty}\frac{1+4c}{k\delta^{2}}\\
 & \leq\frac{1+4c}{\left(N_{n}-1\right)\delta^{2}}
\end{align*}
As $\lim_{n\rightarrow\infty}N_{n}=\infty$, this probability vanishes for sufficiently large values of $n$.

Let us now consider case that $F_{n}$ does not occur. In this case, $\left|U_{n,k^{2}}-\mathbb{E}\left[U_{n,k^{2}}\right]\right|\leq\delta/2$ for each $k\in\left\{ \left\lfloor \sqrt{N_{n}}\right\rfloor ^{2},\left\lfloor \sqrt{N_{n}}\right\rfloor ^{2}+1,\ldots,\left\lfloor \sqrt{n}\right\rfloor^{2}\right\}.$ Applying \cref{lem:closelymatching}, below, it follows that $\left|U_{n,k}-\mathbb{E}\left[U_{n,k}\right]\right|\leq\delta$ for each $k\geq N_{n}$ as long as $N_{n}$ is sufficiently large. On the other hand, by construction, $\mathbb{E}\left[U_{n,N_{n}}\right]\leq\beta q-\delta$, and we thus conclude that $k_{n}^{*}=\max\left\{ k:\ U_{n,k}\leq\beta q\right\} $ must satisfy $k_{n}^{*}\geq N_{n}$.  As $k_n^*$ is itself random, to ensure unambiguous notation, we now define $\mu_{n,k} = \mathbb{E}[U_{n,k}]$.  In this notation, assuming that $F_n$ does not occur, we find that $\left|U_{n,k_{n}^{*}}-\mu_{n,k^*_n}\right|<\delta$ and so $\mu_{n,k^*_n}\leq\delta+\beta q$. Thus, applying Theorem 1, $V_{n}$ is at most $\delta/q+\beta$.  As we have already shown that the probability of $F_n$ not occuring vanishes, it follows that $\lim_{n\rightarrow\infty}\mathbb{P}\left(V_{n}>\beta+\delta/q\right)=0$, as desired. 
\end{proof}

\begin{lemma} \label{lem:closelymatching}
Let $a_{k}=\frac{1}{k}\sum_{i=1}^{k}x_{i}$ and $b_{k}=\frac{1}{k}\sum_{i=1}^{k}y_{i}$, with $x_{i},y_{i}\in\left\{ 0,1\right\} $. Assume $\left|a_{m^{2}}-b_{m^{2}}\right|<\delta$ for all sufficiently large values of $m$. Then $\left|a_{k}-b_{k}\right|=\delta+O(1/k)$.
\end{lemma}

\begin{proof}
Pick $m,k$ with $m^{2}<k<\left(m+1\right)^{2}.$ Then 
\begin{align*}
\left|a_{k}-a_{m^{2}}\right| & =\frac{1}{k}\sum_{i=1}^{m^{2}}x_{i}+\frac{1}{k}\sum_{i=m^{2}+1}^{k}x_{i}-\frac{1}{m^{2}}\sum_{i=1}^{m^{2}}x_{i}\\
 & =\frac{1}{k}\sum_{i=m^{2}+1}^{k}x_{i}+\left(\frac{1}{k}-\frac{1}{m^{2}}\right)\sum_{i=1}^{m^{2}}x_{i}\\
 & \leq\frac{4m+2}{m^{2}}=O(1/m)
\end{align*}
By the same argument argument, $\left|b_{k}-b_{m^{2}}\right|=O(1/m)$. For sufficiently large values of $k$, it follows that $\left|a_{k}-b_{k}\right|\leq\delta+O\left(1/\left\lfloor k^{2}\right\rfloor \right)$, as desired.
\end{proof}

\section{Simulation algorithm}

\label{apdx:simprocess}

Here we describe our algorithm for simulating high-throughput cell perturbation experiments.  The algorithm depends on seven simulation parameters: the number of perturbations ($P$), the number of genes ($G$), the number of independent gene programs ($K$), the degree of influence of those gene programs ($\sigma_1^2$), the level of measurement noise ($\sigma_2^2$), the degrees of freedom of the measurement noise distribution (df), and the number of replicates ($R$).

\begin{algorithm}
\caption{Simulation of a High-Throughput Cell Perturbation Experiment }
\label{alg:simIIIIII}
\begin{algorithmic}[1]  
\Procedure{sim}{$P,G,K,\sigma_1,\sigma_2,\mathrm{df},R$}
    \State $\theta_{n,g} \sim \mathcal{N}(0,(p/P)^2)\quad  \forall p\in\{1,\ldots, P\},g\in \{1,\ldots,G\}$
    \State $A_{g,k} \sim \mathcal{N}(0,1/K) \quad \forall g\in \{1,\ldots,G\}, k\in \{1,\ldots,K\}$ 
    \State $B_{r,p,k} \sim \mathcal{N}(0,1) \quad \forall r\in \{1,\ldots,R\},p\in \{1,\ldots,P\}, k\in \{1,\ldots,K\}$
    \For{$r\in \{1,\ldots,R\},p\in \{1,\ldots,P\}, g\in \{1,\ldots,G\}$}
    \State $C_{r,n,g} = \sum_{k=1}^K A_{g,k}B_{r,p,k}$
    \State $D_{r,n,g} \sim \mathrm{StudentT}(\mathrm{df})$
    \State $X_{r,n,g} = \theta_{n,g} + \sigma_1 C_{r,n,g} + \sigma_2 D_{r,n,g}\sqrt{\frac{\mathrm{df}-2}{\mathrm{df}}}$
    \EndFor
    \State \Return $(\theta, X)$
\EndProcedure
\end{algorithmic}
\end{algorithm}

\section{Dataset acquisition and pre-processing}

\label{apdx:datasets}

We used two datasets, which we refer to as {\LonekA} and {\LonekB}. The {\LonekA} dataset was originally presented by \citet{subramanian2017next} and deposited to the GEO database \citep{L1000}.  It can be viewed at several levels of pre-processing; we elected to view after it had been preprocessed to what is referred to as level 4.  The data in this level comprises a $z$-score for each replicate for each perturbation for each dosage for each cell type for each gene.  The data can be downloaded from GEO accession GSE70138 (cf.~\texttt{https://www.ncbi.nlm.nih.gov/geo/query/acc.cgi}).  

This dataset was reanalyzed by \citet{qiu2020bayesian}, leading to an alternative dataset, also containing $z$-scores, which we refer to as {\LonekB}. This data is available from the following URL: 
\texttt{https://github.com/njpipeorgan/L1000-bayesian}.

Both datasets include information about 978 genes for many different perturbations applied to many different cell types at many different dosages.  We focus on the parameters concerning A375 cells when subjected to the highest dosages considered for each perturbation. 

\section{Discoveries per perturbation}
\label{apdx:L1000discbypert}

\begin{figure}
\begin{center}
\includegraphics[width=\figwidth]{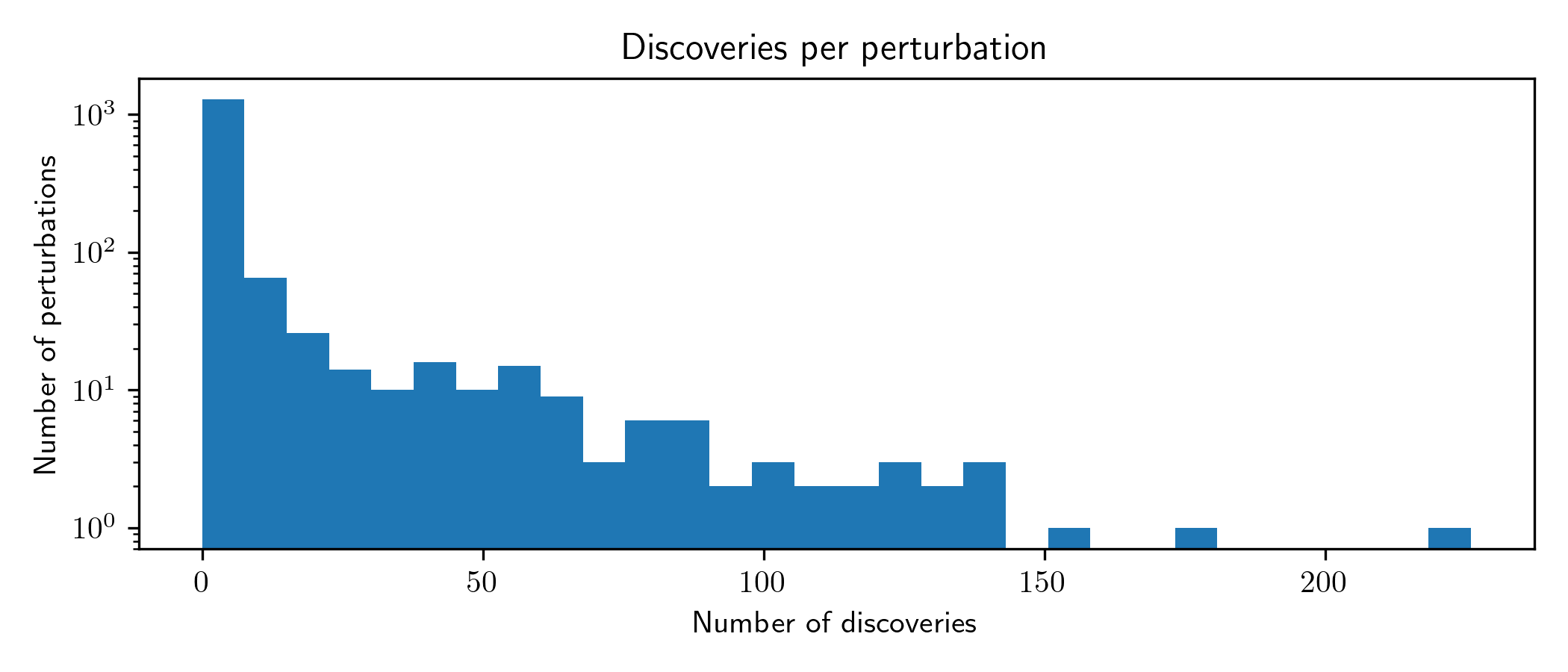}
  \caption{
  \captionheading{Discoveries by perturbations.}  We apply SSE from \cref{sec:method} to L1000 data to select parameters in which the sign can be reliably inferred.  For each perturbation $p$, we then count the number of parameters pertinent to $p$ that were selected.  The figure above shows a histogram of these counts.
  }\label{fig:L1000discbypert}
\end{center}
\end{figure}

We used data from the L1000 platform to consider the effects of 1482 perturbations on 978 genes.  Let $\theta_{p,g}$ denote the parameter that indicates the effect of perturbation $p$ on gene $g$.  The SSE method applied with target type S proportion of 10\% makes 38,283 discoveries based on this data, that is, it selects 38,283 parameters in which the sign can be reliably inferred.  However, these discoveries are not equally distributed among the perturbations.  In fact, for 25\% of the perturbations $p$, there is no gene $g$ such that $\theta_{p,g}$ was selected.  That is, for 25\% of the perturbations, there were no genes for which the sign of effect for that perturbation could be reliably inferred.  \cref{fig:L1000discbypert} shows a complete histogram, indicating the heavy tails of the distribution on genes selected per perturbation.  This figure suggests that there are many perturbations that have little effect and a few perturbations that affect many genes.

\section{The difficulty of applying mirror statistic methods for sign estimation}

\label{apdx:mirrorproblems}

Here we explore whether the method of \citet{dai2023false} could be used to select parameters for the purpose of type S error control.  We present a simple counterexample in which it would be inadvisable to (i) select parameters based on \citet{dai2023false} and then (ii) estimate the parameter's sign for each selected parameter.

Consider a scenario in which half the parameters are $-1$ and half are $+1$.  For each parameter $\theta_i$, we have two independent measurements, $X_{i}^{(1)}$ and $X_{i}^{(2)}$, each drawn from the distribution
\[
\frac{1}{10}\delta_{10}+\frac{9}{10}\delta_{\theta_{i}}.
\] 
That is, $\mathbb{P}\left(X_{i}^{(r)}=\theta_{i}\right)=0.9$ and $\mathbb{P}\left(X_{i}^{(r)}=10\right)=0.1$.  Despite the anomalous value 10, the estimates are faithful.  

We now apply the mirror statistic method of \citet{dai2023false} to select a subset of the parameters.  The assumptions of this method are trivially satisfied, as there are no parameters for which the null hypothesis holds.  Let 
\[
M_{i}=\mathrm{sign}\left(X_{i}^{(1)}X_{i}^{(2)}\right)\left(\left|X_{i}^{(1)}\right|+\left|X_{i}^{(2)}\right|\right)
\]
and consider the sets
\[
\hat{S}=\left\{ i:\ M_{i}\geq100\right\} \qquad\mathrm{and}\qquad\tilde{S}=\left\{ i:\ M_{i}\leq-100\right\}.
\]
The set $\tilde{S}$ is always empty (if the signs disagree, only one of them may have returned the anomalous value). Thus, $\left|\tilde{S}\right|/\left|\hat{S}\right|\vee1=0$.  The mirror statistic method of \citet{dai2023false} will therefore estimate the false discovery proportion of this subset as $\widehat{\mathrm{FDP}}=0$. Thus, the mirror statistic method will select all parameters in $\hat{S}$ for any user-supplied value $q$.

The mirror statistic method works as intended in this case.  It successfully identifies that $\hat S$ contains no parameters in which $\theta_i=0$.  

However, this method does \emph{not} identify a subset of parameters in which the sign of each parameter can be reliably inferred.  In fact, the sign of each parameter in $\hat S$ cannot be inferred from the measurements of these parameters.  For example, if we estimate the signs of these parameters using $X_{i}^{(1)}$ for $i \in S$, then our type S error rate would be 50\%.

It is possible that this method could be guaranteed to control the type S error under further assumptions about the distribution of the measurements (for non-null parameters).  However, it is not immediately clear how this could be done: intuitively, the difficulty is that this method uses both $\boldsymbol{X}^{(1)}$ and $\boldsymbol{X}^{(2)}$ to sort the parameters.  Thus, if there are large anomalous values, the sort order will prioritize the parameters associated with the rare event that the \emph{both} measurements of it were anomalous.

\end{document}